\documentclass[%
 aip,
 amsmath,amssymb,
 reprint,%
]{revtex4-1}

\usepackage{graphicx}
\usepackage{bm}
\usepackage[utf8]{inputenc}
\usepackage[T1]{fontenc}
\usepackage{mathptmx}
\usepackage{xcolor}

\begin{document}

\title{Ultrafast electron dynamics in platinum and gold thin films driven by optical and terahertz fields}

\author{V. Unikandanunni}
\affiliation{Department of Physics, Stockholm University, SE-10691 Stockholm, Sweden}
\author{M.C. Hoffmann}
\affiliation{Linear Coherent Light Source, SLAC National Accelerator Laboratory, 94025 Menlo Park, CA, USA}
\author{P. Vavassori}
\affiliation{CIC nanoGUNE BRTA, San Sebastian, and IKERBASQUE, Basque Foundation for Science, Bilbao, Spain}
\author{S. Urazhdin}
\affiliation{Department of Physics, Emory University, Atlanta, GA, USA}
\author{S. Bonetti}
\email{stefano.bonetti@fysik.su.se}
\affiliation{Department of Physics, Stockholm University, SE-10691 Stockholm, Sweden}
\affiliation{Department of Molecular Sciences and Nanosystems, Ca’ Foscari University of Venice, 30172 Venice, Italy}

\begin{abstract}
We investigate the ultrafast electron dynamics triggered by terahertz and optical pulses in thin platinum and gold films by probing their transient optical reflectivity. The response of the platinum film to an intense terahertz pulse is similar to the optically-induced dynamics of both films and can be described by a two-temperature model. Surprisingly, gold can exhibit a much smaller terahertz pulse-induced reflectivity change and with opposite sign. For platinum, we estimate a 20\% larger electron-phonon coupling for the terahertz-driven dynamics compared to the optically-induced one, which we ascribe to an additional nonthermal electron-phonon coupling contribution. We explain the remarkable response of gold to terahertz radiation with the field emission of electrons due the Fowler-Nordheim tunneling process, in samples with thickness below the structural percolation threshold where near-field enhancement is possible. Our results provide a fundamental insight into the ultrafast processes relevant to modern electro- and magneto-optical applications.
\end{abstract}

\maketitle

The advent of intense single-cycle terahertz (THz) radiation sources has opened new avenues for the exploration of light-matter interaction, thanks to the large accessible electric fields of the order of MV/cm at photon energies in the meV range. With the emergence of intense table-top THz sources, it is now possible to study, in a laboratory setting, THz driven ultrafast dynamics of coupled degrees of freedom in condensed matter systems \cite{vicario2014gv,ruchert2012scaling,liu2012terahertz,kampfrath2013resonant,hafez2016intense}. Efforts underway are aiming at reaching local electric field strengths of the order of 100 MV/cm (1 V/\AA), comparable to that of the interatomic fields using near-field enhancement in metamaterials\cite{salen2019matter}. Metals provide an excellent medium to investigate THz-induced coupled many-body dynamics, thanks to the presence of multiple degrees of freedom able to interact with THz radiation, including free electrons, phonons, and magnons in magnetic materials. A variety of THz-driven physical processes ranging from ultrafast demagnetization in magnetic thin films \cite{bonetti2016thz} to field emission in nanotips have been recently reported in the literature \cite{bonetti2016thz,li2016high}, which were earlier on observed only in the optical regime. \cite{herink2012field,beaurepaire1996ultrafast}. Now, thanks to the availability of intense THz sources, it is possible to compare the dynamics induced by light pulses in two regions of electromagnetic spectrum that differ in frequency by three orders of magnitude. This is expected to provide an unprecedented insight into phenomena involving light-matter interaction \cite{levchuk2020coherent}.

Optically-induced ultrafast dynamics in metals is usually discussed using a phenomenological two-temperature model (2TM) \cite{anisimov1974electron}. This model is in principle applicable to dynamics driven by THz radiation as well, since it does not rely on the specific mechanism of excitation but only on the total energy deposited. According to the 2TM model, ultrafast pulses directly excite energetic electrons, which within tens of femtoseconds thermalize via electron-electron collisions to a Fermi-Dirac distribution at a higher temperature, and only at later times (hundreds of femtoseconds) they thermalize with the lattice. However, time-resolved photoemission spectroscopy measurements have shown that the 2TM assumption of distinct time scales is not always valid, and that a nonequilibrium electron population can exist for up to 1 ps for materials such as gold \cite{fann1992direct,fann1992electron}. No data exist for the case of THz-driven excitations. Even more advanced descriptions, such as the one offered by the Fermi liquid theory, which correctly describe eV excitations with the characteristic $\tau^{-1}\sim(E-E_F)^2$ dependence, break down when the photon energy is lowered towards the meV range. \cite{pines1966theory,kim1992carrier}. In fact, in this energy range, more refined models of ultrafast electron dynamics must also include a temperature dependent term which prevents an indefinite slowdown of the relaxation with lowering photon energy \cite{groeneveld1995femtosecond}. 

\begin{figure}[t]
\includegraphics[width=\columnwidth,scale=1]{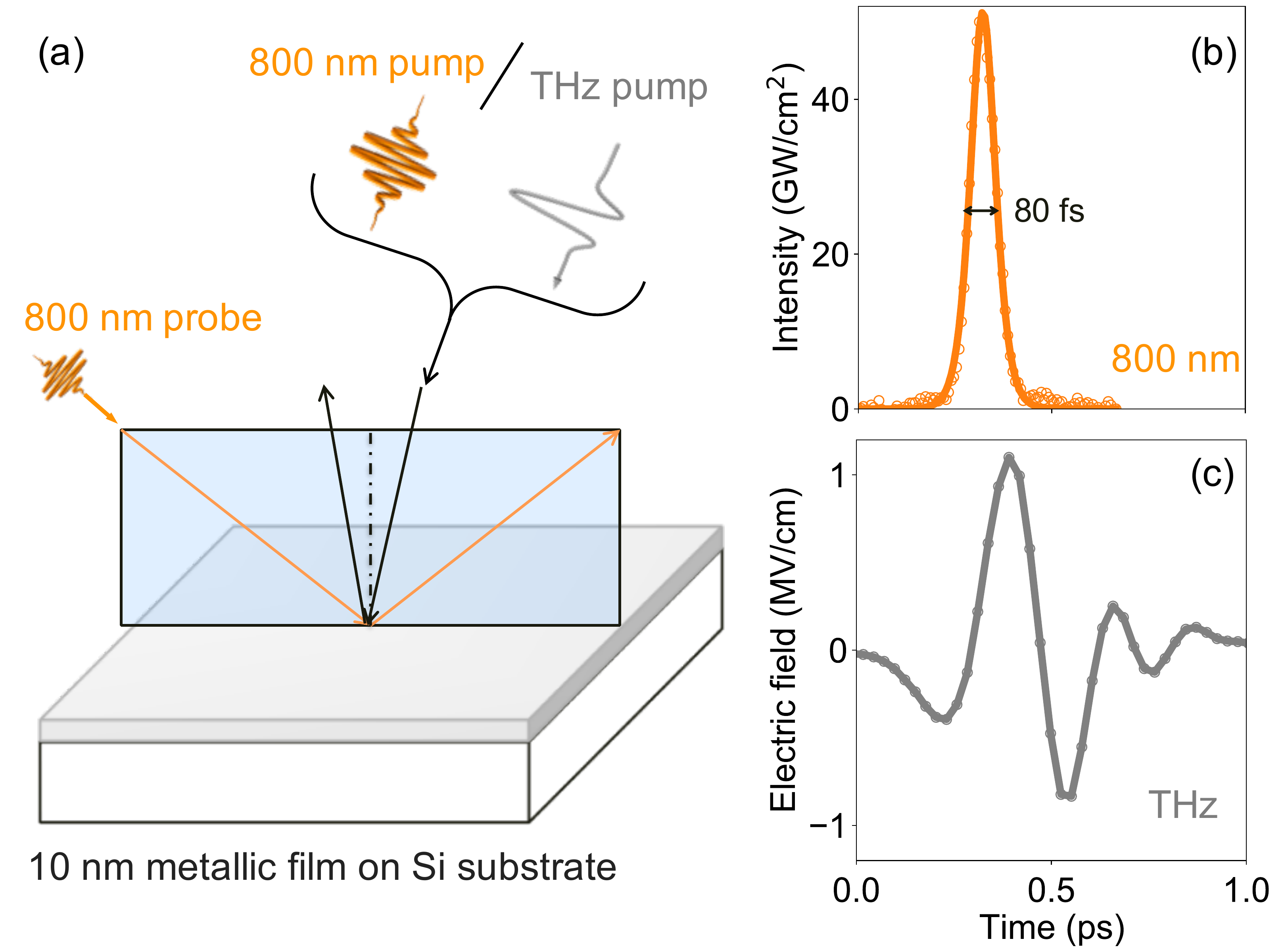}
\caption{\label{fig:1}(a) Geometry of the experimental setup for measuring transient reflectivity. The plane of incidence is marked in blue. (b) Autocorrelator trace of the 800 nm pump intensity. (c) Temporal profile of the electric field of THz pulse. The symbols in panels (b) and (c) are the measured data. The solid line in panel (b) is the best fit obtained with a Gaussian function, the line in panel (c) connects the data points. } 
\end{figure}

In this Letter, we compare the dynamics induced by THz and near-infrared radiation in 10-nm-thick platinum and gold films, two metals that are neighbors in the periodic table but have very different band structures. Both materials are of a broad current interest in condensed matter physics \cite{liu2021differentiating,wilson2020parametric,zahn2020ultrafast}. Platinum is widely used in magnetic and magneto-optical experiments and is a key material in spintronics due to its large spin-orbit coupling, whereas gold is essential in plasmonics and for the fabrication of metamaterials thanks to its low ohmic losses. By performing time-resolved reflectivity measurements using 800 nm probe pulses, and by simulating the dynamics using a 2TM, we gain a detailed understanding of the dynamics at play, which we expect to be also relevant to other metals with similar electronic configuration.

\begin{figure}
\includegraphics[width=\columnwidth]{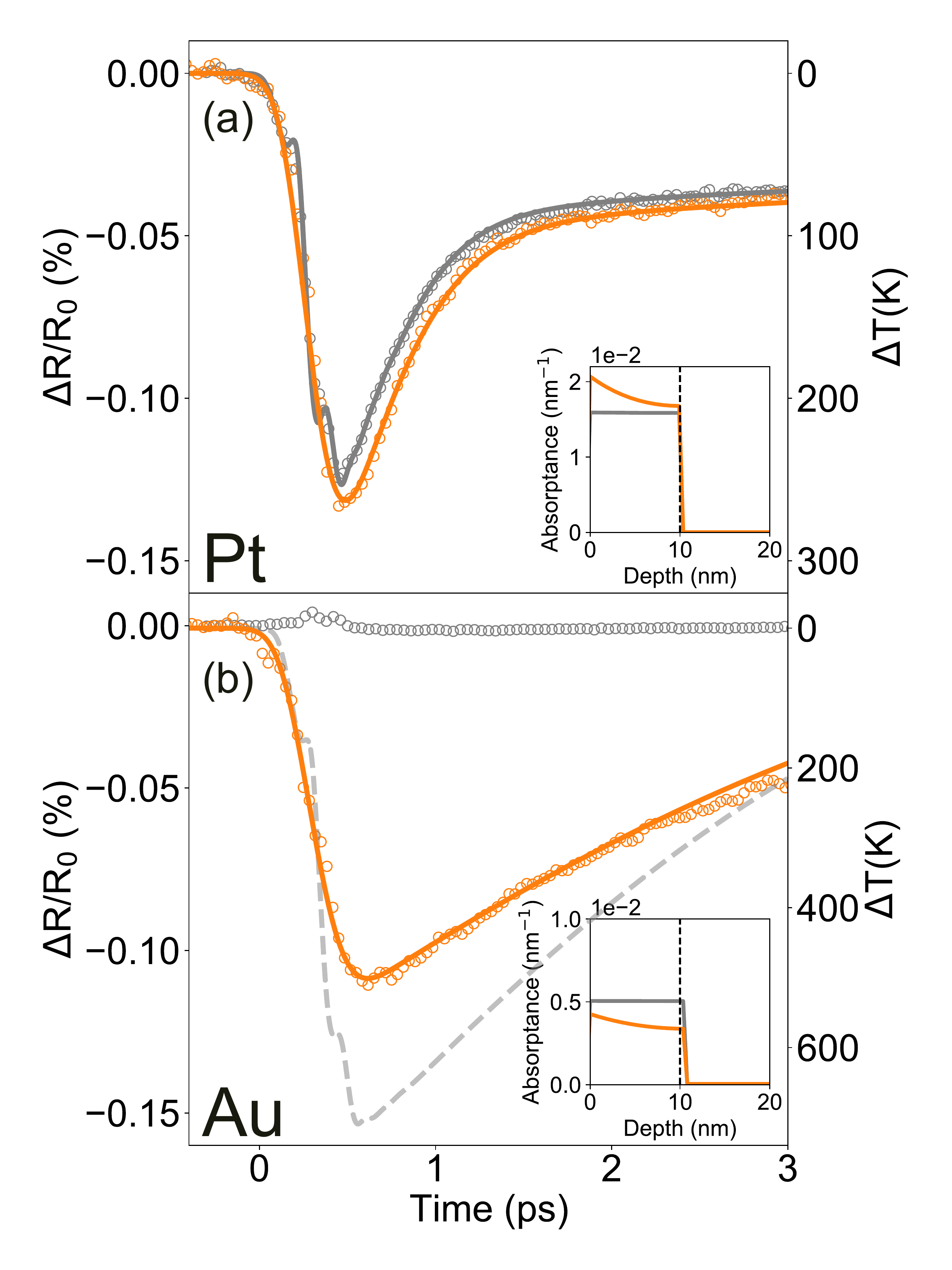}
\caption{\label{fig:2}Transient reflectivity of (a) platinum and (b) gold samples after excitation with THz (gray) and 800 nm (orange) pump pulses of comparable fluence of approx. 2.5 mJ/cm$^2$. The solid and dashed lines show the temperature evolution calculated using the 2TM model described in the text, with the temperature scale shown on the right axis. The insets represent the simulated absorption profile in the for the two pump wavelengths.}
\end{figure}

The 800 nm pump-probe experiments are performed with an amplified Ti:sapphire laser, with a pulse duration of approximately 55 fs and a repetition rate of 1 kHz. We choose orthogonal polarization between the pump (s-polarized) and the probe (p-polarized) beams to eliminate the coherent artifacts which arise in the interference between energy-degenerate pulses. The pump is incident at an angle $\theta_{pump}$ = 10 degrees and the probe at an angle $\theta_{probe}$ = 45 degrees, as shown in Fig. \ref{fig:1}(a). The temporal profile of the pulse intensity measured using an autocorrelator is shown in Fig. \ref{fig:1}(b).

The intense single-cycle THz pump required for the THz pump-optical probe experiment is generated in the organic crystal DSTMS by the optical rectification of 1560 nm radiation, down-converted from the 800 nm fundamental using an optical parametric amplifier \cite{vicario2014generation}. The temporal profile of the electric field of the THz pump, determined by electro-optic sampling in a GaP crystal \cite{nahata1996coherent}, has a peak amplitude of approximately 1 MV/cm, as shown in Fig. \ref{fig:1}(c). The incident fluence is around 2.5 mJ/cm${^2}$ for both THz and 800 nm pumps. The change in reflectivity of the 800 nm probe is measured using a silicon photo-detector. For both experiments, the pump is modulated with a mechanical chopper at half the laser repetition rate.

Fig. \ref{fig:2}(a)-(b) shows the transient reflectivity in platinum and, respectively, gold. For both materials, the reflectivity change caused by the THz pump is shown by gray symbols, while that caused by the 800 nm pump by orange ones. The continuous lines show the 2TM simulation results. The calculated absorbance of both pumps as a function of sample depth is given in the insets. Both calculations are based on the transfer matrix approach, implemented using the open-source simulation package NTMpy \cite{alber2020ntmpy}. The material parameters and other simulation details are provided in the Supplementary Material.

\begin{table}[b]
\caption{\label{tab:1}Electron-phonon coupling $G$ extracted from the data in Fig. \ref{fig:2} using the 2TM simulation, and the corresponding values from the literature \cite{smirnov2020copper,caffrey2005thin,lin2008electron,hohlfeld2000electron}.}
\begin{ruledtabular}
\begin{tabular}{ccc|c}
 &$G_{\rm THz}$ (W/m$^3$K) &$G_{800}$ (W/m$^3$K) &$G^{\rm Ref. \cite{smirnov2020copper,caffrey2005thin,lin2008electron,hohlfeld2000electron}}$ (W/m$^3$K)\\
\hline
Platinum&(11.2$\pm$0.1)$\times$10$^{17}$&(9.3$\pm$0.1)$\times$10$^{17}$&(2.5-11)$\times$10$^{17}$\\
Gold& - & (2.2$\pm$0.1)$\times$10$^{16}$&(2.1$\pm$0.3)$\times$10$^{16}$
\end{tabular}
\end{ruledtabular}
\end{table}

We first discuss the results for platinum in Fig.~\ref{fig:2}(a). The transient reflectivity shows a qualitatively similar response for both THz pump and 800 nm pumps, with a maximum relative change reaching approximately 0.1\% after less than 1 ps. After that, the reflectivity rapidly relaxes to within 0.05\% after a couple of picoseconds, followed by a slow recovery towards equilibrium at longer times. The absorbed fluence is similar for the two different pump energies, and it leads to a maximum temperature change of approximately 250 K according to the 2TM simulations. These simulations also allows to extract the electron-phonon coupling $G$ for the two cases, as shown in the first row of Table \ref{tab:1}. We note that the electron-phonon coupling extracted for the THz-induced dynamics $G^{\rm Pt}_{\rm THz}$ is about 20\% larger than the one for optical frequencies $G^{\rm Pt}_{800}$. This is a relevant finding, since the electron-phonon coupling is usually assumed to be independent of the pump wavelength, at least in the optical region.

Moving to gold, Fig. \ref{fig:2}(b), for the 800 nm pump, the initial transient decrease of reflectivity is slightly slower than for platinum, but still occurs within less than 1 ps, and reaches a similar maximum relative variation of 0.1\%. The subsequent recovery towards equilibrium is much slower than for platinum; the  electron-phonon coupling extracted from these data using the 2TM simulations is indeed 50 times smaller. The same model returns a maximum temperature increase of approximately 450 K. Aside from these quantitative differences, the transient reflectivity with the 800 nm pump is qualitatively similar for the two metals. On the contrary, the reflectivity change in gols driven by the THz pump is much different: it is more than an order of magnitude smaller in amplitude, and opposite in sign. Before going into a detailed explanation, we point out that this is not a trivial effect due to a smaller pump absorption. The inset of Fig. \ref{fig:2}(b) shows that, on the contrary, the absorbance of the THz pump is comparable and larger than for the 800 nm light.

\begin{figure}
\includegraphics[width=1\columnwidth]{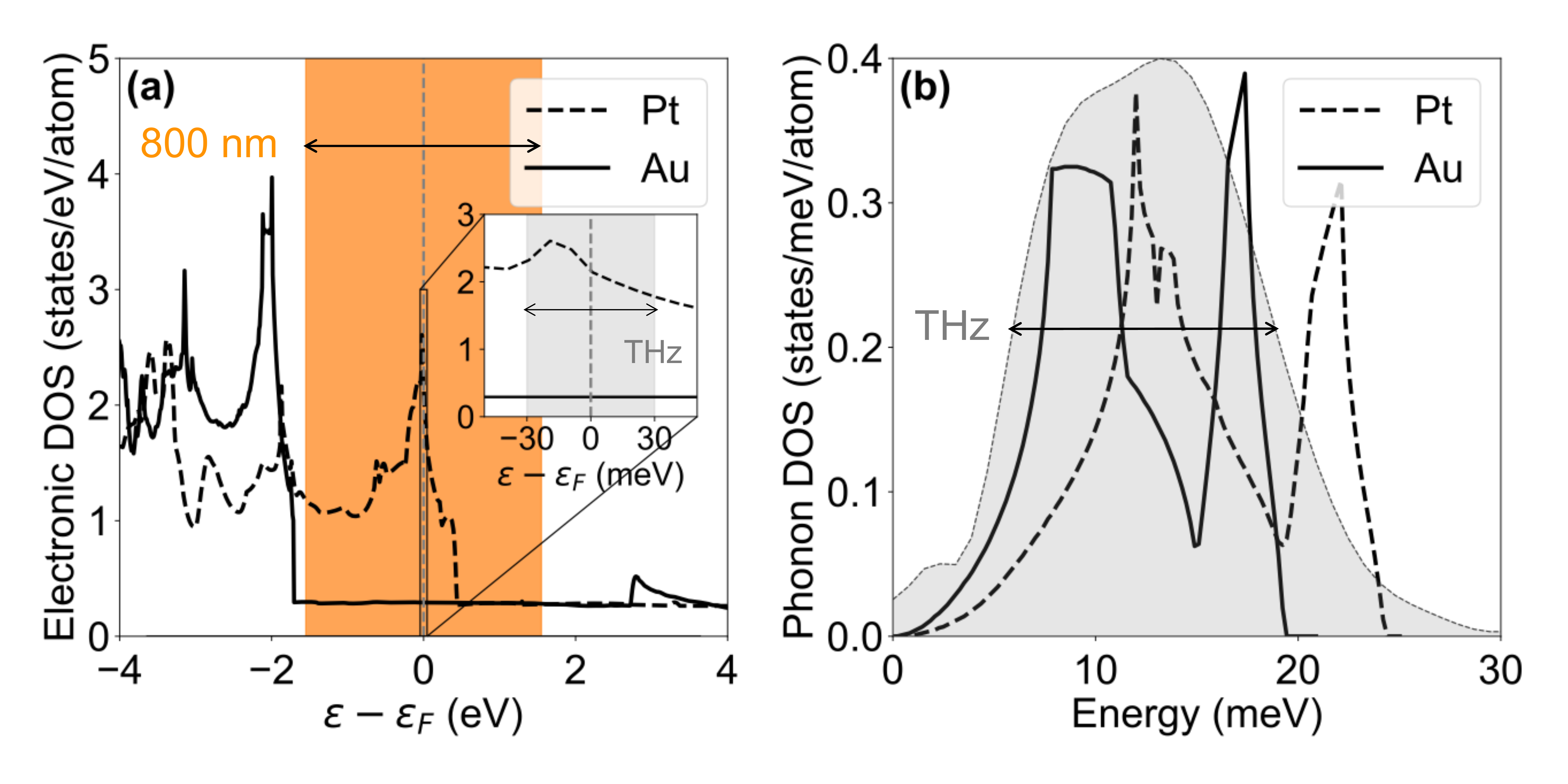}
\caption{\label{fig:3}(a) Electronic and (b) phonon density of states (DOS) for platinum and gold. The calculated DOS are taken from Refs.  \cite{lin2008electron,schober1981phonon,zhalko1985calculation,hellwege1981metals}. The orange and gray shading illustrate the maximum bandwidth of the nonthermal electrons excited by the 800 nm and, respectively, THz pumps.}
\end{figure}

To gain insight into the mechanisms determining the transient reflectivity, we plot in Fig. \ref{fig:3}(a)-(b) the electronic and the phonon density of states (DOS) for the two materials. The orange and gray shading represent the initial bandwidth of the non-thermal electrons excited by the 800 nm pump and, respectively, the THz pump. The 800 nm pump excites electrons up to 1.55 eV away from the Fermi level, whereas the THz pump up to a maximum of 30 meV. It is known that electrons located close to the Fermi level are those contributing the most to the electron-phonon coupling \cite{kaganov1957relaxation,ginzburg1955electron}. For the 800 nm excitation, the majority of the nonthermal electrons are far from the Fermi level, so the contribution of electron-phonon coupling to their relaxation is initially negligible, and becomes effective only after they thermalize \cite{fann1992electron}.

We apply these general observations to  first to the case of platinum, where the two excitations give rise to similar dynamics. For both the 800 nm and THz pumps, a relatively large fraction of electrons excited by the electromagnetic field have $d$ character, which have a relatively large scattering cross sections with phonons. For the case of terahertz fields, the characteristic energy of the excited electrons is lower, and consequently the contribution of electron-phonon scattering is larger \cite{bonetti2016thz,levchuk2020coherent}, explaining the observed $G^{\rm Pt}_{\rm THz}>G^{\rm Pt}_{\rm 800}$.

For the case of gold, the situation is remarkably different. The phonon DOS is similar to that of  platinum. However, the electronic DOS is substantially different. In particular, the density of states within about $1$ eV from the Fermi level is dominated by itinerant $s$-states, with a negligible contribution from the localized $d$-states. The $s$-character of the conduction electronic states, as well as the much smaller DOS around Fermi level of gold, result in a much weaker electron-phonon coupling consistent with the observed G$^{\rm Au}_{800}\ll G^{\rm Pt}_{800}$ (see Table 1). However, this observation by itself does not provide an explanation for neither the much weaker THz-induced reflectivity change, nor its opposite sign compared to the 800 nm pump, which is the most striking result in Fig. \ref{fig:2} and in clear contrast with the 2TM calculations. We note in particular that for thermalized electrons, a THz-induced increase of the transient reflectivity would imply a decrease of temperature, in an apparent violation of the conservation of energy. 2TM simulations predict for Au a comparable electronic temperature change for both optical and THz pumps (Fig. \ref{fig:2}(b)), confirming the breakdown of the model, and suggesting that a mechanism other than thermalization must also be at play in the Au film excited by THz fields.

We argue that our observations relate to the emission from the gold film of electrons accelerated by the strong THz electric field. Qualitatively, this mechanism is understood with considering the THz field tilting the potential barrier, enabling electron tunneling into vacuum without sample heating. Electron emission has already been observed in nano-tips and metallic metasurfaces made of gold, tungsten and other materials \cite{lange2020ultrafast,cocker2013ultrafast,li2016high}, where the near-field enhancement helps reaching the threshold of emission. More recently, THz-driven electron emission was observed from a gold surface using a peak electric field as low as 50 kV/cm without any local enhancement, although the thickness and other film characteristics were not reported \cite{li2019terahertz,lombosi2015thz}. The observation of field emission in ultrathin gold films deposited on a solid substrate is however still missing, a geometry that is the most relevant for many applications.

Field-induced electron emission is explained by the Fowler-Nordheim model of quantum mechanical tunneling \cite{fowler1928electron}. The probability of field emission for THz electric field of frequency \textit{f} and peak amplitude $E_{THz}$ is related to the value of Keldysh parameter\cite{keldysh1965ionization} $\gamma_k = 2\pi f \sqrt{2m\phi}/eE_{THz}$, where $\phi$ $m$, and $e$ are the the work function, the mass and, respectively, the charge of the electron. Field emission is dominant for $\gamma_k\ll1$, whereas photoionization for $\gamma_k$ $\gg$ 1. For our experimental parameters and $\phi=5.2$ eV, $\gamma_k \approx 1.2$. When $\gamma_k\approx1$ which physical effect would be dominant is less clear. However, Uiberacker et al \cite{uiberacker2007attosecond} have reported field emission for a value of Keldysh parameter as high as three.

\begin{figure}
\includegraphics[width=\columnwidth]{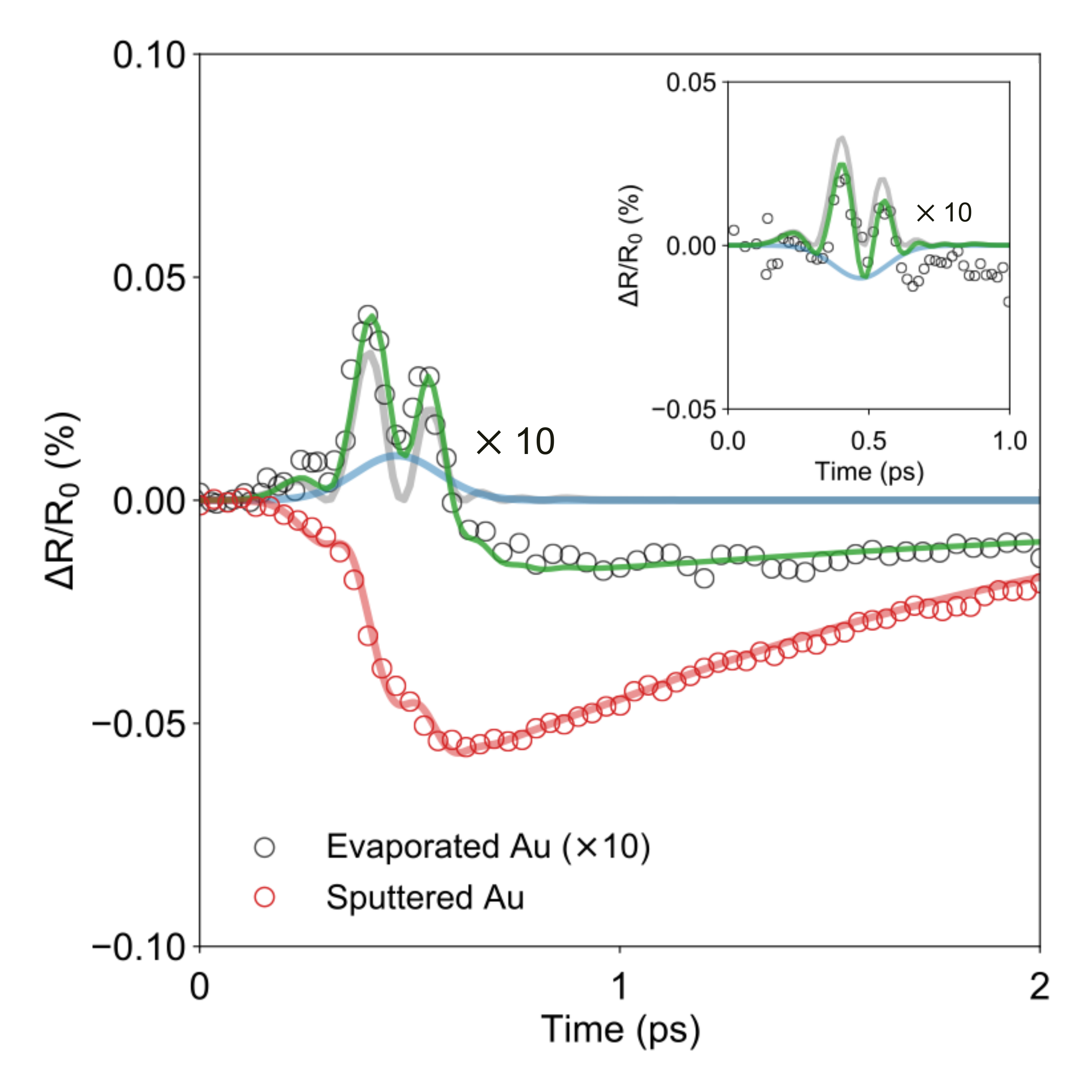}
\caption{\label{fig:4}Symbols: Experimental transient reflectivity of gold in the first $2$~ps excited by the terahertz field $E_{\rm THz}$. Gray solid line: $E_{\rm THz}^2$, with $E_{\rm THz}$ measured independently using electro-optic sampling in a GaP crystal. Blue solid line: THz field rectification modelled as a scaled Gaussian envelope of $E_{\rm THz}$. Green solid line: fit of the data including adding up $E_{\rm THz}^2$, rectification signal and an exponential recovery. Inset: same as the main panel, but with the THz pump field polarity reversed by 180 degrees.}
\end{figure}

To further test our hypothesis, in Fig.~\ref{fig:4} we zoom in on the initial transient response of gold (evaporated) to the THz field, together with the response of a sputtered gold film. By overlaying the square of the measured THz electric field $E_{\rm THz}$ on the evaporated gold data, it is clear that the time profile of the initial transient reflectivity goes as $E_{\rm THz}^2$. Since the Fowler-Nordheim tunneling current is also proportional to $E_{\rm THz}^2$ \cite{fowler1928electron}, this further supports the argument that the THz induced transient reflectivity in gold is dominated by electron field emission. The sign of the effect, i.e. the transient \emph{increase} of the reflectivity in the first 500 fs, is expected when the plasma frequency decreases. This is in turn compatible with a Drude-Lorentz picture when the electron density reduces, as it would be the case when electron field emission occurs. The rather laborious algebraic derivation is presented in the Supplementary Material.

In addition, we also observe a rectification behavior, modeled with a Gaussian-like shape mimicking the terahertz field carrier envelope. Such behavior flips its sign when we change the polarization of the THz pump pulse by 180 degrees, as shown in the inset of Fig.~\ref{fig:4}. This rectification effect is thought as the net normal component of the THz electric field arising due to slight asymmetry of the film interfaces (air/gold and gold/substrate). The subsequent decrease of reflectivity at later times is understood as the thermalization of electrons the with the lattice, observed in all other measurements shown in Fig. \ref{fig:2}. However, the effect is now significantly reduced, since most of the energy from the radiation has already been transferred into kinetic energy of the emitted electrons which have left the sample. A good fit to the data is obtained when the three effects: electron field emission, rectification and thermalization are considered. Crucially, we notice that all these effects are gone and the conventional 2TM behavior is recovered in the sputtered film.

Field emission is expected to have negligible amplitude for pump wavelengths in the visible and near-infrared range. Indeed, for visible photons, $\gamma_k$ is two orders of magnitude larger than for the terahertz case. The field emission probability decreases also with the character of the accelerated electrons. Theoretically, the tunneling probability for $d$-electrons is approximately three orders of magnitude lower than that of $s$-electron with the same energy \cite{gadzuk1969band}. This is due to the larger scattering cross section of $d$-electrons which prevents them to be accelerated at large enough energies, scattering and thermalizing with the lattice before. Pt has a relatively large density of $d$-states around the Fermi level, as shown in Fig. \ref{fig:3}(a), and hence electron field emission is not expected in this material. Finally, the 2TM-like behavior in the sputtered gold film with much lower roughness than the conventional evaporated one, suggest that the microscopic structure of the gold film plays a key role in the process, since nm-scale inhomogeneity leads to near-field enhancement of the THz radiation. This lowers $\gamma_k$ further into the field emission regime. We stress that the nanometer roughness is a well-known characteristic of thin gold films below a critical thickness of the order of 10 nm, not a sign of lower quality samples. The films are homogeneous at spatial scales comparable with the wavelength of visible and terahertz light, and the transient reflectivity induced by 800 nm light has the expected response for gold, as shown in Fig. \ref{fig:2}(b). We show in the Supplementary Material that a thicker gold film (50 nm, well above the percolation threshold) deposited by evaporation, show no measurable field emission nor negative transient reflectivity. This indicates that both the electric field enhancement due to inhomogeneity and the temperature rise are greatly suppressed when the film is homogeneous and thick enough, and that all the effects observed in our work are relevant for films with thicknesses of 10 nm or less.

In summary, we studied the THz and 800 nm induced electron and lattice dynamics in gold and platinum thin films measuring their transient optical reflectivity. Platinum showed comparable response to both THz and optical pumps, with a 2TM simulation returning a 20\% larger electron-phonon coupling $G$ when the material excited by terahertz fields. We suggest that this can be understood introducing an additional coupling channel only available when nonthermal electrons are present close to the Fermi level, which is the case for terahertz radiation but not for near-infrared light. Hence, our results suggest that the common assumption of wavelength-independent $G$ is not strictly correct when one considers wavelengths which differs by orders of magnitude. Gold, when driven with 800 nm pump light, also showed a response consistent with the 2TM. However, we found that such model breaks down when the material is excited with intense terahertz fields in evaporated films below the percolation threshold. We argue that such behavior can be explained in terms of electronic field emission within the Fowler-Nordheim model, supported by a reasonably small Keldysh parameter and a positive transient reflectivity which varies as the square of the incident terahertz field. The much larger fraction of $s$-states near the Fermi level in gold, as compared to platinum, makes this mechanism the dominant effect in this case, further promoted by the local electric field enhancement caused by the nm-scale inhomogeneity of thin gold films. We anticipate that our results will be highly relevant for many ultrafast experiments in magnetism and optics where platinum, gold or metallic thin films with similar properties are incorporated as a part of heterostructures or metamaterials.

See the supplementary material for (i) two-temperature model simulations; (ii) sample characterization using THz time-domain spectroscopy; (iii) derivation of the relationship between transient optical reflectivity and material properties; and (iv) THz transient reflectivity in 50 nm films.

V.U. and S.B acknowledge support from the European Research Council, Starting Grant 715452 “MAGNETIC-SPEED-LIMIT”, S.U. acknowledges support from the National Science Foundation Award Nos. ECCS-1804198 and ECCS-2005786. P.V. acknowledges support from the Spanish Ministry of Economy, Industry and Competitiveness under the Maria de Maeztu Units of Excellence Programme - MDM-2016-0618.

\section*{Data availability}
The data that support the findings of this study are available from the corresponding author upon reasonable request.

\bibliographystyle{apsrev}
\providecommand{\noopsort}[1]{}\providecommand{\singleletter}[1]{#1}

\end{document}